\documentclass[12pt,preprint]{aastex}

\def\masyr{{\rm mas}\,{\rm yr}^{-1}}
\def\usno{{\rm USNO}}
\def\nltt{{\rm NLTT}}

\def\lim{{\rm lim}}

\begin{document}

\title{Revised NLTT}

\author{Samir Salim and Andrew Gould}
\affil{Department of Astronomy, The Ohio State University,
140 W.\ 18th Ave., Columbus, OH 43210}
\email{samir,gould@astronomy.ohio-state.edu}

\singlespace

\begin{abstract}
	
    Revised NLTT (New Luyten Catalogue of Stars With Proper Motions Larger
than Two Tenths of an Arcsecond) contains improved astrometry and new
optical/infrared photometry for the vast majority of NLTT stars lying in the
overlap of regions covered by POSS I and by the 2MASS release, approximately
44\% of the sky.  The epoch 2000 positions are typically accurate to 130 mas,
the proper motions to $5.5\,\masyr$, and the $V-J$ colors to 0.25 mag.
Relative proper motions of binary components are meaured to $3\,\masyr$.  The
false identification rate is $\sim 1\%$ for $11\la V\la 18$ and substantially
less at brighter magnitudes.

\end{abstract}
\clearpage

\section{{Description of the Revised Catalog}
\label{sec:catalog}}

	Construction of the revised catalog and the properties of the original and the revised NLTT are discussed in \citet{paper1,paper2}.

	The revised catalog contains information grouped in six sections,
1) summary, 2) NLTT, 3) source identifications, 4) USNO, 5) 2MASS,
6) binaries, plus the last column which is an internal reference for
debugging.

\subsection{{Summary Information}
\label{sec:suminfo}}

This section contains 11 entries: 1) the NLTT number (drawn consecutively
from 1 to 58,845), 2) a letter code `A', `B', or `C' if the NLTT ``star''
has been resolved into several sources, 3) $\alpha$ (2000, epoch and equinox), 
4) $\delta$ (2000),
5) $\mu_\alpha$, 6) $\mu_\delta$, 7) $\sigma(\mu_\alpha)$, 8) 
$\sigma(\mu_\delta)$ (all four in arcsec yr$^{-1}$), 9) $V$, 10) $V-J$,
11) 3-digit source code.

The three digits of the source code refer to the sources of the position,
proper motion, and $V$ photometry.  1 = Hipparcos, 2 = Tycho-2, 
3 = Tycho Double Star Catalog (TDSC), 4 = Starnet,
5 = USNO/2MASS, 6 = NLTT, 7 = USNO (for position) or common proper motion 
companion (for proper motion).  More specifically, ``555'' means
2MASS based position, USNO based $V$ photometry, and USNO/2MASS based
proper motion.

	The (2000) position has been evolved forward from whatever epoch
it was measured using the adopted proper motion.  When the positon source
is Hipparcos, Tycho-2, or TDSC, the position is given in degrees 
to 6 digits, otherwise to 5 digits.  
For proper motions derived from a PPM catalog, the errors are adopted
from that catalog.  Proper motions from USNO/2MASS determinations 
have an estimated error of $5.5\,\masyr$.  NLTT proper motions are $20\,\masyr$.
CPM binary companions (without other astrometry) are not assigned an error,
and zeros are entered into the error fields.  

As described in Paper I, $V$ refers to the Johnson $V$ entry for Hipparcos,
Tycho $V$ for Tycho-2 and TDSC, and Guide-star catalog $R$ for Starnet.
The conversion from USNO photometry is given by,
\begin{equation}
\label{eqn:usnov}
      V = R_\usno + 0.23 + 0.32(B-R)_\usno.
\end{equation}
We remind the reader that for USNO-A1 photometry, we first convert to
USNO-A2 using equation
\begin{equation}
     B_{\rm A2} = B_{\rm A1} + 49.056 - 9.5613 B_{\rm A1} 
+0.669 B_{\rm A1}^2 -0.0198 B_{\rm A1}^3
+0.0002 B_{\rm A1}^4,\qquad (B_{\rm A1} > 13.07)
\label{eqn:rusno12}
\end{equation}
 before applying equation
(\ref{eqn:usnov}).  When NLTT photometry is used, $V$ is evaluated using
equations:
\begin{equation}
R_\usno = 0.9333 R_\nltt + 0.6932,\qquad (R_\nltt>10.39)
\label{eqn:rusnonltt}
\end{equation}
and $R_\usno = R_\nltt$ for $R_\nltt<10.39$, and 
$$(B-R)_\nltt - (B-R)_\usno = 0.2387 - 0.055 R_\nltt,\quad (R_\nltt <16.5)$$
\begin{equation}
(B-R)_\nltt - (B-R)_\usno = 7.3817 + 0.4067 R_\nltt,\quad (R_\nltt \geq16.5).
\label{eqn:brusnonltt}
\end{equation}
 to convert
to USNO mags, and then applying equation (\ref{eqn:usnov}).  In the 
rare cases for which NLTT photometry is employed and one of the two
bands is not reported, a color of $(B-R)_\nltt = 1$ is assumed.
No effort has been made to ``de-combine'' photometry in the case of unresolved
binaries.  For example, if a binary is resolved into two stars in
2MASS, but is unresolved in USNO, then different $J$ band measurements
will be reported for the two stars, but both with have the same,
combined-light $V$ photometry.  The $V-J$ color reported in field 10
will be the simple difference of these two numbers.  Similarly, if
the NLTT ``star'' is resolved by TDSC but not 2MASS, then the
$V$ light will be partitioned between the two stars but not the $J$
light.  Finally, note that if no $J$ band photometry is available
(whether because of saturation, faintness, or the star being in an area
outside the second incremental 2MASS release) $V-J$ is given as $-9.$

When multiple sources of information are available, the priority for what
is presented in the summary is as follows. 
Positions: 3,1,2,5,4,7,6;  
Proper Motions: 3,2,1,5,4,7,6; 
Photometry: 3,1,2,5,4,6.

As discussed in Paper I, Tycho-2 proper motions are given precedence over
Hipparcos primarily because they better reflect the long-term motion
when the stars are affected by internal binary motions, but also because
at faint magnitudes they are generally
more precise.  When Hipparcos proper motions are given, it is often because
the star is so faint that it does not show up in Tycho.  In this case,
the nominal Hipparcos errors are often quite large and true errors can
be even larger.  We found a handful of cases by chance in which the Hipparcos
proper motion was grossly in error and we removed the Hipparcos entry
and substituted the USNO/2MASS value.  However, we made no systematic
effort to identify bad Hipparcos proper motions.

Only stars for which we are providing additional information are recorded
in the catalog.  There are 36,020 entries for 35,662 NLTT stars including
a total of 723 entries for 361 NLTT ``stars'' that have been resolved in TDSC.

\subsection{{NLTT Information}
\label{sec:nlttinfo}}

The next 6 columns give information taken from NLTT, namely
12) $\alpha$ (2000), 13) $\delta$ (2000), 14) $\mu_\alpha$, 15) $\mu_\delta$,
16) $B_\nltt$, 17) $R_\nltt$.  The coordinates and proper motions are
precessed from the original 1950 equinox to 2000, and the position is updated
to 2000 epoch using the NLTT proper motion.

\subsection{{Source Information}
\label{sec:sourceinfo}}

The next 2 columns give source information.  Column 18 is the Hipparcos number
(0 if not in Hipparcos).  Column 19 is the identifier from TDSC, Tycho-2,
or Starnet, whichever was used to determine the position in columns 3 and 4.
When the position comes from Hipparcos, 2MASS, or USNO, or NLTT, ``null'' is
entered in this field with one exception: when a Starnet measurement
has been superseded 
by a 2MASS measurement, the Starnet identifier has been retained for ease
of recovery of this source.  It can easily be determined that the summary
information comes from 2MASS because the first digit in field 11 
will be a ``5''.

\subsection{{USNO Information}
\label{sec:usnoinfo}}

The next six fields give USNO information:
20) Integer RA, 21) Integer DEC, 22) $B_{\rm A1\ or\ A2}$, 
23) $R_{\rm A1\ or\ A2}$, 
24) USNO Epoch,
25) 3-digit search-history code.  The Integer RA and DEC together serve
as a unique USNO identifier since that is the form RA and DEC are given in
the original USNO-A1 and USNO-A2 releases.  They can also be converted into 
degree $\alpha$ and $\delta$ (at the USNO epoch) using the formulae: 
$\alpha$ = (Integer RA)/360000, $\delta$ = (Integer DEC)$/360000 - 90$. 
Regarding the 3-digit search history code, the first
digit tells which USNO catalog the entry is from: 1 = USNO-A1, 2 = USNO-A2.
The second tells whether the USNO source was found in the rectangle (1) or
the circle (2).  The third tells whether it was a unique match (1), or had
to be resolved by hand from among several possible matches (2).
If there is no USNO information, all of these fields are set to zero.

\subsection{{2MASS Information}
\label{sec:2massinfo}}

The next six fields contain 2MASS information:
26) $\alpha$, 27) $\delta$ (both at 2MASS Epoch), 28) $J$, 29) $H$,
30) $K_s$, 31) 2MASS Epoch.  If no 2MASS data are available, all fields
are replaced by zeros.  If there are 2MASS data, but not for a particular
magnitude measurement, that value is replaced by $-9$.

\subsection{{Binary Information}
\label{sec:binaryinfo}}

The next six fields contain information about binarity:
31) binarity indicator, 32) NLTT number of binary companion, 33) NLTT
estimated separation, 34) NLTT estimated position angle, 35) our
estimated separation, 36) our estimated position angle.  Regarding
the binarity indicator, 0 means NLTT does not regard this as a binary.
Otherwise, it is a NLTT binary and the 
indicator is set according to whether the companion
is (2) or is not (1) in our catalog.  The NLTT estimates of the separation
position angle come from the NLTT Notes.  Our estimates come from the
difference of the 2000 positions of the two stars.  In cases for which
the companion is not in our catalog, the fields with ``our'' separation
and position angle are replaced by values found from the difference of
the NLTT coordinates (i.e., fields 12 and 13). The companion numbers are based on what we think is
obviously what Luyten intended, rather than what was literally written
down.  However, no effort has been made to clean up any other transcription
errors, even when these are equally obvious.  No binary information is
recorded in these fields about NLTT ``stars'' that were resolved by TDSC.
Rather, the reader should recognize each of those binaries from the upper case
letter appended to its NLTT number (column 2).

The final column (37) is an integer that is used in the program that
assembles the catalog from the various subcatalogs that are described in
Paper I and II.  It is useful mainly to us, but for completeness:
1 = PPM+2MASS, 2 = USNO/2MASS, 3 = CPM, 4 (not used), 5 = 2MASS-only,
6 = PPM (no 2MASS), 7 = Annulus match, 8 = USNO-only.  

The catalog is available from 
\url{http://www.astronomy.ohio-state.edu/\~{}gould/NLTT}.
\notetoeditor{Previously ApJ had problems with interpreting urls that 
contain tilde. For reference, the address should have a slash, then tilde 
in front of gould. No spaces} 
The Fortran format statement for the catalog record is:\hfil\break\noindent
(i5,a1,2f11.6,2f8.4,2f7.4,2f6.2,1x,3i1,2f10.5,2f8.4,2f5.1,i7,1x,a12,
\hfil\break\noindent
2i10,2f5.1,f9.3,i2,2i1,2f10.5,3f7.3,f9.3,i2,i6,f7.1,f6.1,f7.1,f6.1,i2)


\begin{thebibliography}{}

\bibitem[Gould \& Salim(2002)]{paper1} Gould, A.\ \& Salim, S.\ 2002
\apj, accepted, astro-ph/0204217

\bibitem[Salim \& Gould(2002)]{paper2} Salim, S.\ \& Gould, A.\ 2002
\apj, submitted, astro-ph/0206318


\end{thebibliography}
\end{document}